\documentclass{JAC2003}

\usepackage{graphicx}

\begin{document}
\title{UNAVERAGED THREE-DIMENISONAL MODELLING OF THE FEL}

\author{C.K.W. Nam, P. Aitken, and B.W.J. McNeil,\\
Department of Physics, University of Strathclyde, Glasgow, G4 0NG, Scotland, UK}

\maketitle

\begin{abstract}
A new three-dimensional model of the FEL is presented. A system of
scaled, coupled Maxwell--Lorentz equations are derived in the
paraxial limit. A minimal number of limiting assumptions are made
and the equations are not averaged in the longitudinal direction
of common radiation/electron beam propagation, allowing the
effects of coherent spontaneous emission and non-localised
electron propagation to be modelled. The equations are solved
numerically using a parallel Fourier split-step method.
\end{abstract}

\section{INTRODUCTION}

A three-dimensional model of a helical wiggler Free Electron Laser
(FEL) is presented that minimises the assumptions made.  This
model does not average the Maxwell--Lorentz equations describing
the interaction between the electrons, wiggler and radiation
fields.  Furthermore, no relativistic approximations in the
equations governing electron motion are made and transverse motion
of the electrons is self-consistently driven by both the wiggler
and radiation fields. Current 3D codes perform averaging over a
radiation wavelength of the radiation and others also over the the
electron trajectories. Thus no study of the effects of coherent
spontaneous emission (CSE) or self amplified coherent spontaneous emission
(SACSE) is possible. Furthermore, these codes cannot model large
energy losses by the electrons as the electrons are confined
locally to a wavelength.  In this model no such averaging is done
and so sub-period radiation evolution is included and electron
migration over distances greater than a wavelength, for example
due to locally high radiation powers, may be described.

\section{3D MODEL}
Following the method of previous studies for the one-dimensional
FEL \cite{{cse-ref},{helicalwiggler-ref},{shotnoise-ref}} the
physics of the FEL may be described in three dimensions by the
coupled Maxwell--Lorentz equations
\begin{equation}\label{eq:MaxwellEqn}
 \left(\nabla^2-\frac{1}{c^2} \frac{\partial^2}{\partial t^2}\right)
 \mathbf{E}(\mathbf{r},t)= \mu_0 \frac{\partial \mathbf{J}_\perp(\mathbf{r},t)}{\partial t}
\end{equation}
\begin{equation}\label{eq:LorentzEqn}
\frac{d\mathbf{p}_j}{dt} =
-e\left(\mathbf{E}(\mathbf{r}_j,t)+\frac{\mathbf{p}_j}{\gamma_j m}
\times \mathbf{B}(\mathbf{r}_j,t) \right)
\end{equation}
where $j=1\ldots N$, the total number of electrons, and the
transverse current density may be written
\begin{equation}\label{eq:Current}
\mathbf{J}_\perp(\mathbf{r},t)= -\frac{e}{m} \sum_{j=1}^N
\frac{\bar{p}_{\perp
j}}{\gamma_j}\delta(\mathbf{\bar{r}}-\mathbf{\bar{r}}_j(t)),
\end{equation}
where $m$ is the electron rest mass.  The helical wiggler and
radiation electric field are assumed to be
\begin{eqnarray}
{\bf B}_w(\mathbf{r}) &=& \frac{B_w}{\sqrt{2}} (\hat{\bf{e}}\,e^{-ik_w z}+c.c.)\\
& & \hspace{-1cm}+ \frac{B_w k_w}{2}[i x (e^{-ik_w z}-c.c.)+y(e^{ik_w z}+c.c.)]\hat{\bf{z}} \nonumber\\
{\bf E}(\mathbf{r},t) &=&\frac{1}{\sqrt{2}} (\hat{\bf{e}}\;
\xi_0({\bf r},t) e^{i(kz-\omega t)} + c.c. )
\end{eqnarray}
where $\hat{\bf{e}}= (\hat{\bf{x}} + i\hat{\bf{y}})/\sqrt2$,
 $\xi_0$ is a slowly varying complex field envelope and $B_w$ is
the wiggler magnetic field strength of period $\lambda_w =
2\pi/k_w$. The magnetic component of the radiation field
$\mathbf{B}(\mathbf{r},t)$ is required in the Lorentz equation and
may be calculated from Maxwell's equations to give:
\begin{equation}
\mathbf{B}(\mathbf{r},t) = -\frac{i}{\sqrt{2}} \left(\frac{\xi_0({\bf r},t)}{c}
\;\hat{\bf{e}}\, e^{i(kz-\omega t)} - c.c. \right)
\end{equation}
where it has been assumed that any radiation that is
counter-propagating the electron beam may be neglected. The
following scaling notation is now introduced:
\begin{eqnarray*}
\varepsilon Q_j &=& \frac{1-\beta_{z_j}}{\beta_{z_j}},
\varepsilon = \frac{1-\bar{\beta}_z}{\bar{\beta}_z},
\alpha=\left(\frac{2\rho\gamma_r}{a_w} \right)^2, \nonumber\\
\bar{p}_\perp &=& \frac{p_x-ip_y}{mc}, A = \frac{e\xi_0}{mc \omega_p\sqrt{\gamma_r \rho}}, \nonumber \\
\bar{z} & = & 2 k_w \rho z, \ \bar{z}_2 = 2 k_w \rho \frac{\bar{\beta}_z}{1-\bar{\beta}_z}(ct-z), \\
    \bar{x} & = & \frac{x}{\sqrt{l_g l_c}},\  \bar{y} = \frac{y}{\sqrt{l_g l_c}}
\end{eqnarray*}
where $\bar{\beta}_z = v_{z0}/c$ is the mean electron velocity on
entering the interaction region, $\gamma_r$ is the resonant
electron energy and $\rho$ is the fundamental FEL parameter,
defined as
\[
\rho = \frac{1}{\gamma_r} \left( \frac{a_w \omega_p}{4ck_w}\right)^{2/3}.
\]
Here $\omega_p = \sqrt{e^2 n_p/(\varepsilon_0 m)}$ is the plasma
frequency for the peak electron number density of the electron
pulse $n_p$, and $a_w=eB_w/(mck_w)$ is the wiggler deflection
parameter.  The scaled electron density is defined as $\bar{n}_p =
n_p l_g l_c^2$, where $l_g$ is the gain length and $l_c$ the
cooperation length.  Note the scaling in the transverse plane is
with respect to the physically relevant gain and cooperation
lengths. This allows the equations above describing the FEL
interaction to be written as
\begin{eqnarray}\label{eq:waveEqn}
 \lefteqn{-i\rho \left(\frac{\partial^2 A}{\partial \bar{x}^2}+
\frac{\partial^2 A}{\partial \bar{y}^2}\right)+
\frac{\partial A}{\partial \bar{z}}+2i \rho \frac{\partial^2
A}{\partial \bar{z}\partial \bar{z}_2}  =}   \\
\nonumber & & \frac{\gamma_r}{a_w} \frac{1}{\bar{n}_p}
\sum_{j=1}^N \frac{\bar{p}_{\perp j}(\varepsilon Q_j(\varepsilon
Q_j+2))^{1/2} }{(1+|\bar{p}_{\perp j}|^2)^{1/2}}
\delta(\mathbf{\bar{r}}-\mathbf{\bar{r}_j})
e^{i\frac{\bar{z}_{2j}}{2\rho}}
\end{eqnarray}
\begin{eqnarray}
\label{eq:pPerpEqn}
\frac{d \bar{p}_{\perp j}}{d \bar{z}} & = &
\frac{a_w}{2\rho} \left[ie^{-i\frac{\bar{z}}{2\rho}} -
\alpha \varepsilon Q_j A
e^{-i\frac{\bar{z}_{2j}}{2\rho}} \right] - \\
&  & \mbox{} \frac{a_w^2
\varepsilon}{8\rho^2}\sqrt{\frac{Q_j(2+\varepsilon
Q_j)}{(1+|\bar{p}_{\perp j}|^2)}} \frac{(\bar{x}_j -
i\bar{y}_j)}{(1+\varepsilon Q_j)^2}\nonumber
\\
\label{eq:QEqn}
\nonumber
\frac{d Q_j}{d\bar{z}} & = &\frac{a_w}{4\rho} \frac{Q_j
(\varepsilon Q_j +2)}{1+|\bar{p}_{\perp j}|^2} \times \\
\nonumber
& & \left[i(\varepsilon Q_j+1)(\bar{p}_{\perp j}^{*}
e^{-i\frac{\bar{z}}{2\rho}} - c.c.) + \right.\\
& & \left.\alpha\varepsilon Q_j (\bar{p}_{\perp j}^{*}  A
e^{-i\frac{\bar{z}_{2j}}{2\rho}} + c.c.) \right]
\\
\label{eq:z2Eqn}
\frac{d\bar{z}_{2j}}{d\bar{z}} &=&  Q_j
\\
\label{eq:xEqn}
\frac{d\bar{x}_j}{d\bar{z}} &=&
\Re(\bar{p}_{\perp j})\sqrt{\frac{Q_j(2+\varepsilon Q_j)}{1+|\bar{p}_{\perp j}|^2}}\\
\label{eq:yEqn}
\frac{d\bar{y}_j}{d\bar{z}}  &=&
-\Im(\bar{p}_{\perp j})\sqrt{\frac{Q_j(2+\varepsilon Q_j)}{1+|\bar{p}_{\perp j}|^2}}
\end{eqnarray}
In deriving the equations the only approximations made are the
neglect of space charge and the paraxial approximation.  There are
no restrictions on the electron energy allowing large changes to
be modelled.

In the scaled variables used here, the normalised beam emittance is given by
\begin{equation}
    \epsilon_n = \sqrt{l_g l_c} \sigma_{\bar{x}}\sigma_{\bar{p}_x}.
\label{eq:epn}
\end{equation}
For an electron beam matched to a focussing channel the radius of
the electron beam envelope is given by
\begin{equation}
 r_{b}=\left(\frac{\epsilon_n\beta}{\gamma_r}\right)^{1/2}.
\label{eq:MatchedR}
\end{equation}
The electrons experience a natural focussing force due to the
magnetic field of the helical wiggler \cite{greenbook-ref}
resulting in a beta-function of $\beta=\gamma_r/a_wk_w$. This
restoring force is approximated by the final  linear term in
(\ref{eq:pPerpEqn}) and may be simply modified in magnitude to
approximate other, e.g.\ FODO, focussing channels.

The scaled Rayleigh range (in units of $\bar{z}$) can be shown to
be $\bar{z}_R=\sigma_{\bar{r}}^2/2\rho$, where $\sigma_{\bar{r}}$
is the radius of the beam in scaled units of $\bar{x},\bar{y}$.

It can be shown that the above equations satisfy energy
conservation, written in the form:
\begin{eqnarray}
\label{energy} \underbrace{\int\int\int}_{all space}|A|^2
d\bar{x}\;d\bar{y}\;d\bar{z}_2 +\frac{1}{\gamma_r
\rho}\sum_{k=1}^{N_p} \bar{\chi}_k \gamma_k=0 \\
\mbox{ where } \gamma_k = \sqrt{\frac{(1+|\bar{p}_{\perp
k}|^2)(1+\varepsilon Q_k)^2}{\varepsilon Q_k(\varepsilon Q_k+2)}}
\nonumber
\end{eqnarray}
which may be used as a check in the numerical solution.

In the relativistic Compton limit $\epsilon,\rho\ll 1$ the above
equations can be shown to reduce to those of~\cite{cse-ref}.

\section{NUMERICAL SOLUTION}
The field evolution of the scaled equations
(\ref{eq:waveEqn})--(\ref{eq:yEqn}) are solved by applying a
modified parallel Fourier split-step method \cite{FFT-ref} in
conjunction with a finite element method \cite{FEM-ref}. This
allows the effects of coherent spontaneous emission (CSE),
self-amplified spontaneous emission (SASE) and diffraction to be
modelled numerically.  The parallel code was developed using the
Numerical Algorithms Group (NAG) parallel routines to control the
parallel processing. The Fourier split-step method involves
solving the wave equation in two steps. The first step considers
the field diffracting in the transverse direction and propagating
in the $\bar{z}$ direction freely without the electron source
term:
\begin{equation}\label{eq:waveEqnSplitStep1}
-i\rho \left(\frac{\partial^2 A}{\partial \bar{x}^2}+
\frac{\partial^2 A}{\partial \bar{y}^2}\right)+
\frac{\partial A}{\partial \bar{z}}+2i \rho \frac{\partial^2
A}{\partial \bar{z}\partial \bar{z}_2}  = 0.
\end{equation}
This equation can be solved by taking the Fourier transforms in
$\bar{x}$, $\bar{y}$ and $\bar{z}_2$ resulting in an ordinary
differential equation in the Fourier transformed field
$\tilde{A}$, with the following analytic solution:
\begin{equation}
\label{eq:ODESolution} \tilde{A}(\bar{z}+\Delta\bar{z}) =
\tilde{A}(\bar{z}) \exp\left[\frac{-i\rho k_\perp^2}{(1-2\rho
k_{\bar{z}_2})}\Delta\bar{z}\right],
\end{equation}
where the transverse Fourier transform variable pairs are given by
$(\bar{x},k_{\bar{x}})$,  $(\bar{y},k_{\bar{y}})$,
$(\bar{z}_2,k_{\bar{z}_2})$ and $k_\perp^2 = k_{\bar{x}}^2 +
k_{\bar{y}}^2$. The inverse numerical Fourier transform is then
applied giving solution
$A(\bar{x},\bar{y},\bar{z}_2,\bar{z}+\Delta\bar{z})$ which is then
used as the initial field for the second part of the split-step
method where the diffraction terms are neglected and the source
term acts alone:
\begin{eqnarray}\label{eq:waveEqnSplitStep2}
& &\frac{\partial A}{\partial \bar{z}}+2i \rho \frac{\partial^2
A}{\partial \bar{z}\partial \bar{z}_2}  = \\
\nonumber& &\frac{\gamma_r}{a_w} \frac{1}{\bar{n}_p} \sum_{j=1}^N
\frac{\bar{p}_{\perp j}(\varepsilon Q_j(\varepsilon Q_j+2))^{1/2}
}{(1+|\bar{p}_{\perp}|^2)_j^{1/2}}
\delta(\mathbf{\bar{r}}-\mathbf{\bar{r}_j})
e^{i\frac{\bar{z}_{2j}}{2\rho}}.
\end{eqnarray}
The finite element method is then used to solve the wave equations
(\ref{eq:waveEqnSplitStep2}) and
(\ref{eq:pPerpEqn})--(\ref{eq:yEqn}) along with a fourth order
Runge-Kutta method for electron variables to give the final
solution.

Before applying the finite element method to the wave equation,
the summation over the real electrons has to be changed to a
summation over a group of macroparticles, each of which
represents many real electrons. This reduces the
computational memory load and operation:
\begin{equation}
\frac{1}{\bar{n}_p} \sum_{j=1}^N (\cdots)_j = \frac{1}{\bar{n}_p}
\sum_{k=1}^{N_p} N_k (\cdots)_k
\end{equation}
where subscripts $j$ and $k$  indicate evaluation at the particle
position, $N_p$ is the total number of macroparticles and $N_k$ is
the charge weight of the macroparticle in units of the electron
charge.

The Galerkin method of the finite elements is used.
The field is described by a set of 8-node hexahedral elements with linear
basis functions $\Lambda_m (\bar{x},\bar{y},\bar{z}_2)$ and nodal values $a_m(\bar{z})$:
\begin{equation}
A(\bar{x},\bar{y},\bar{z},\bar{z}_2)= \sum_m a_m(\bar{z})
\Lambda_m (\bar{x},\bar{y},\bar{z}_2)\label{eq:splitA}.
\end{equation}
where $m$ is the global node index of the 3D system. Note that
$\Lambda_m (\bar{x},\bar{y},\bar{z}_2)\equiv 0$ outwith the
elements to which it belongs. The wave
equation~(\ref{eq:waveEqnSplitStep2}) then reduces to a matrix
equation for the nodal points:
\begin{eqnarray}
{\bf K} \frac{\partial a_m(\bar{z})}{\partial\bar{z}} =
 \frac{\gamma_r}{a_w}  \sum_{k=1}^{N_p}&&
\frac{\bar{\chi}_k}{V_e} \frac{\bar{p}_{\perp k}(\varepsilon Q_k
(\varepsilon Q_k+2))^{1/2}}{(1+|\bar{p}_{\perp}|^2)^{1/2}}\times \nonumber \\
&&\Lambda_m(\bar{x}_k,\bar{y}_k,\bar{z}_{2k})
e^{i\frac{\bar{z}_{2k}}{2\rho}} \nonumber
\end{eqnarray}
where $\bf K$ is the stiffness matrix constructed from the
elemental equations~\cite{FEM-ref}, $\bar{\chi}$ is a
macroparticle weighting function and $V_e$ is the scaled volume of
an element of the system, $V_e=\Delta \bar{x} \Delta \bar{y}
\Delta \bar{z}_2$.

\section{A SIMULATION IN THE 1D LIMIT}
To test the 3D model a comparison with the 1D results
of~\cite{cse-ref,shotnoise-ref} was made.  In these
one-dimensional models the wave equation was integrated
analytically over the common electron/radiation transverse area
before numerical solutions were obtained.

\begin{figure}
\includegraphics[width=85mm]{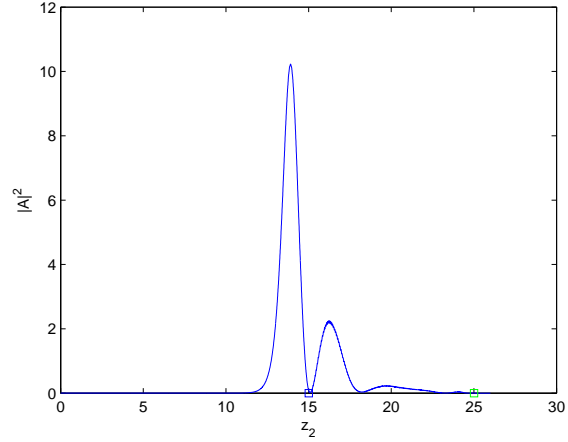}
\caption{\label{fig1}The scaled radiation power $|A|^2$ as a
function of $\bar{z}_2$ demonstrating good agreement with the 1D
model of ~\cite{cse-ref} Fig.~2}
\end{figure}

\begin{figure}
\includegraphics[width=85mm]{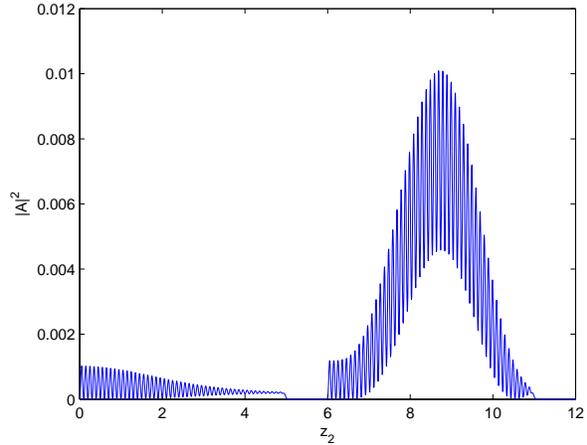}
\caption{\label{fig2}The scaled radiation power $|A|^2$ as a
function of $\bar{z}_2$ demonstrating good agreement with the 1D
model of~\cite{shotnoise-ref}~Fig.~4}
\end{figure}

Here, one element is used in the transverse plane approximating
the radiation field to a plane wave and allowing integration over
the transverse area and comparison with the 1D results. The
electron beam radius was set greater than the electron orbital
radius ensuring the electron beam does not move significantly in
the transverse plane of the element. A gaussian distribution for
the electrons in the transverse plane was used with a range of six
times the standard deviation.  The results of~\cite{cse-ref}
Fig.~2 showing self-amplified coherent spontaneous emission
(SACSE) for a top-hat pulse current are reproduced in
Fig.~\ref{fig1}. (Note that there is a scale reversal
as~\cite{cse-ref} plots the power as a function of $\bar{z}_1$
while here it is plotted as a function of $\bar{z}_2$.)
The coherent spontaneous emission, including the effects of shot
noise, from both square and gaussian shaped pulses
of~\cite{shotnoise-ref} were also reproduced. In Fig.~\ref{fig2}
the top hat case of~\cite{shotnoise-ref}~Fig.~4 is plotted.
Good agreement between the 1D models
of~\cite{cse-ref,shotnoise-ref} and the 1D limit of the 3D model
of this paper are observed.

\section{A SIMULATION IN 3D}
An example of the code operating in 3D is now presented. The
parameters used are not intended to model any specific system, but
rather to demonstrate the functioning of the code and its
associated post-processing and plotting routines. A relatively
simple system was chosen with a  short electron pulse so
that coherent spontaneous effects can be observed and large
computational effort is not required.

The energy spread and the emittance of the beam were set to zero,
the emittance by setting $\sigma_{\bar{p}_{x,y}}=0$ for all
electrons. Focussing was therefore not required for this
simulation. The code was run using 16-processors of a parallel
computer with parameters shown in Table~\ref{table}.
\begin{table}[hbt]
\begin{center}
\begin{tabular}{|l|l|}
\hline
\textbf{Electron beam parameters}& \\
Energy \ E&$250$ MeV\\
Bunch Charge \ Q & $100$ pC\\
Peak Intensity \ $I_{pk}$ & $\sim 9394$A\\
Distribution  in $\bar{x} \ \& \ \bar{y}$ & Gaussian \\
Sigma  in $\bar{x},\bar{y}$ \ $\sigma_{\bar{x},\bar{y}}$ & $\sim 0.228$ \\
Distribution  in $\bar{z}_2$ & Top-hat \\
Length of pulse in $\bar{z}_2$& $1.5$\\
Emittance and energy spread&0\\
\textbf{Undulator parameters}&\\
Undulator Type & Helical\\
Pierce parameter \ $\rho$ & $\sim 1.5$e$-2$ \\
Wiggler deflection parameter \ $a_w$ & $1.5$\\
Propagation distance \ $\bar{z}$ & $5.00$ \\
\textbf{Radiation parameters}&\\
Initial seed field over pulse \ $A_0$ & 0.01 \\
Seed distribution in $\bar{x} \ \& \ \bar{y}$ & Gaussian \\
Sigma in $\bar{x},\bar{y}$ \ $\sigma_{\bar{x},\bar{y}}$ & $\sim 0.228$ \\
Rayleigh length \ $\bar{z}_R$&1.74\\
Distribution of seed in $\bar{z}_2$ & Top-hat \\
Length of seed in $\bar{z}_2$& $1.5$\\
\hline
\end{tabular}
\caption{\label{table}3D parameters}
\end{center}
\end{table}

\begin{figure}
\includegraphics[width=85mm]{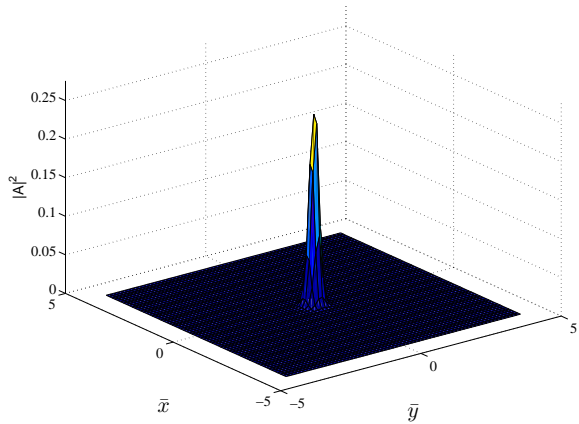}
\caption{\label{fig3}The scaled radiation intensity $|A|^2$ in the
transverse $(\bar{x},\bar{y})$ plane at $\bar{z}=5$. The
transverse slice is taken at the `head' of the electron pulse at
$\bar{z}_2=5$.}
\end{figure}

\begin{figure}
\includegraphics[width=85mm]{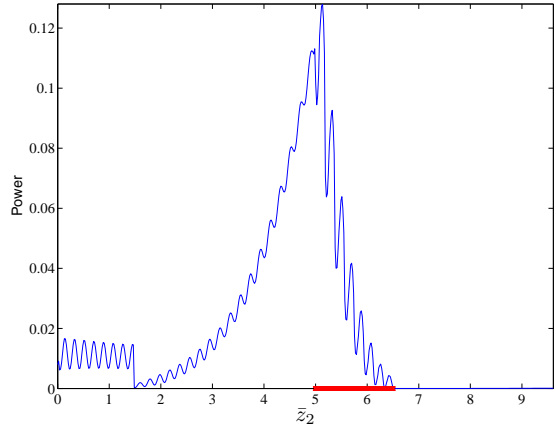}
\caption{\label{fig4}The scaled radiation power at $\bar{z}=5$ as
a function of $\bar{z}_2$. The electron pulse is shown on the
$\bar{z}_2$ axis lying between $5<\bar{z}_2<6.5$. In this frame
the electron pulse propagates along $\bar{z}_2$ with increasing
$\bar{z}$.}
\end{figure}

Fig.~\ref{fig3} plots the scaled intensity $|A|^2$ across a
transverse slice at the head of the electron pulse, just as the
radiation escapes to propagate into vacuum. In Fig.~\ref{fig4} the
scaled power ($|A|^2$ integrated across the transverse plane) is
plotted as a function of $\bar{z}_2$ following a FEL interaction
to $\bar{z}=5$. Both plots demonstrate the main features of FEL
interaction with Coherent Spontaneous Emission and the energy
conservation relation of~(\ref{energy}) is confirmed to within
0.5\%.

Further testing of the code is being carried out and will be
reported upon in a future publication.

\section{CONCLUSION}
A new parallel code has been developed which models the FEL
amplifier by solving a system of scaled equations  describing the
FEL interaction in three spatial and the time dimension. The aim
has been to introduce as few approximations into the model as
possible, the main assumptions being the neglect of space charge
and any backward propagating radiation fields. This allows the
effects of coherent spontaneous emission, diffraction and full
electron transport throughout the region of integration to be
modelled. Furthermore, the sub-wavelength discretisation of the
model allows a significantly wider range of radiation frequencies
to be modelled than is possible with most other codes that use a
minimum discretisation interval of a radiation wavelength. A 1D
limit of the computational model was identified and simulation
results in this limit show agreement with previous 1D numerical
and analytical models. A further example simulation has been
presented which demonstrates the code operating successfully using
an electron pulse in three dimensions. Further optimisation of the
code is ongoing. While the code is undoubtedly slower to operate
than other 3D averaged codes, the extended physics that it can
model may be expected to yield interesting new phenomena in FEL
physics.

\end{document}